**Estimating the volume and surface area of air bubbles entrained by breaking waves from whitecap observations: With implications on the characteristic breaking wave speed and breaking strength parameter**[*]

Paul A. Hwang[1]

Remote Sensing Division, Naval Research Laboratory, Washington DC

[1]*Corresponding author address:* Dr. Paul A. Hwang, Remote Sensing Division, Naval Research Laboratory, 4555 Overlook Avenue SW, Washington DC 20375

Email: paul.hwang@nrl.navy.mil

**ABSTRACT**

A conceptual model relating the whitecap coverage to the bubble plume buoyancy is developed following the observation that the entrained bubble plume buoyancy constitutes a large portion of the breaking wave energy dissipation. The formulation leads to estimations of an effective or equivalent-buoyancy depth of bubble entrainment as well as the volume and surface area of bubbles entrained by surface wave breaking. The results show that the air-water interface area per unit sea surface area is enhanced dramatically by the entrained bubbles: on the order of 10 m$^2$ at about 15 m s$^{-1}$ wind speed. The effective entrainment depth represents the vertical reach of the bubble plume as if all the bubbles were collected into this depth. Based on empirical observations of whitecaps and breaking wave energy dissipation, it is about 0.11 m and relatively independent on wind speed. The void fraction of the top meter ocean layer is related linearly to the whitecap coverage with a proportionality factor of 0.11. The nearly-constant effective entrainment depth essentially renders the bubble entrainment process during the active wave breaking stage into a lateral 2D problem. Published high speed video recordings of bubble plume evolution appear to support this conclusion. Consistent with the nearly-constant effective entrainment depth, relevant breaking wave speeds are within a narrow range between about 2 and 3.5 m s$^{-1}$ and depend on wind speed only weakly. Whitecap observations can also be used to quantify some elusive breaking properties such as the breaking strength parameter *b* relating the breaking energy dissipation rate and length of breaking front.

## 1. Introduction

The phenomenon of surface wave breaking is of great interest to many areas of research, including surface wave dynamics, air-sea interaction, upper ocean dynamics, underwater acoustics, and ocean remote sensing using acoustic and electromagnetic techniques. Among the different methods of measuring wave breaking properties, whitecap observation is probably the most convenient and it has yielded a large volume of data. Analyses of these extensive data sets generally confirm the positive correlation between wave breaking and whitecap coverage $f_w$. Particularly, numerous empirical formulas show that the dependence of $f_w$ on wind speed is cubic, which is a signature of surface wave breaking (Phillips 1985, 1988). For example, a list of 30 formulas is given in Anguelova and Webster (2006), and Goddijn-Murphy et al. (2011) compare the whitecap measurements by Callaghan et al. (2008) with 38 different forms of empirical whitecap equations expressed as functions of wind speed or dissipation rate. There are also several dimensionally-consistent expressions relating $f_w$ to the breaking wave energy dissipation rate (e.g., Toba and Chaen, 1973; Phillips 1985; Toba and Koga 1986; Zhao and Toba 2001; Yuan et al. 2009).

In this paper, a different approach to analyze the whitecap data is proposed for the purpose of extracting additional information beyond confirming the connection between whitecap coverage and wave breaking. The approach is based on the observation that "a large fraction (30-50% and maybe more) of the energy lost [by breaking waves] is entraining the bubble plume" and that "the initial volume of air entrained correlates with energy dissipated" (Lamarre and Melville 1991, p. 471). Section 2 presents a formulation of bubble plume buoyancy $B$ and the rate of energy change $E_{tb}$ expected from the buoyancy. The formulation reveals that, in terms of the bubble properties, the cubic velocity scale in $f_w$ and $E_{tb}$ is the product $gz_ew_b$, where $g$ is the gravitational acceleration, $w_b$ is the representative bubble rise velocity and $z_e$ is an effective or equivalent-buoyancy entrainment depth of the bubble plume, that is, a conceptual depth such that all bubbles in the bubble plume are collected within this depth while maintaining the same ocean surface coverage area. Section 3 presents a connection between $E_{tb}$ and the breaking wave energy dissipation rate $E_{tD}$ derived from the empirical growth function of wind-generated waves (Hwang and Sletten 2008).

As described in section 4, the resulting formulation allows a quantitative evaluation of $z_e$ and other related information, including the volume and surface area of the entrained bubbles by breaking waves. In essence, $z_e$ is proportional to the square root of $E_{tD}/f_w w_b$. To estimate $w_b$ for computing $z_e$, the bubble size distribution of the bubble plume during the active wave breaking stage is an important piece of information. Obtaining the size distribution of bubbles in the ocean is a very difficult task. It is especially challenging for the near surface region and under wave breaking condition. Nonetheless, several papers have reported quantitative results of the near-surface bubble size distribution under breaking waves (Deane 1997; Deane and Stokes 1999, 2002, 2010). The results from field and laboratory measurements show that the peak of the bubble volume size distribution is in the neighborhood of 1 mm radius. Related studies of bubble crushing and fragmentation in turbulent shear flows (Longuet-Higgins 1992; Martínez-Bazán et al. 1999a, b, 2010) also support the robustness on the peak size of the volume distribution in the neighborhood of 1 mm. These studies are described in more detail in the section. Using three sets of empirical functions for $f_w$ and $E_{tD}$, the numerical value of $z_e$ is found to be about 0.11 m and almost independent of wind speed. Two factors contribute to this rather surprising result: (a) the dependence of both $E_{tD}$ and $f_w$ on wind speed is cubic thus canceling out the wind speed factor, and (b) the rise velocity of large bubbles with radius greater than about 0.5 mm is relatively invariable with respect to the bubble size (Gaudin 1957; Clift et al. 1978; Leifer et al. 2000), thus further removing the wind speed factor embedded in the possible wind speed dependence that may exist in the representative bubble size of the entrained bubble plume.

Section 5 presents the computational results of the conceptual model using the whitecap function of Phillips (1985) based on his breaking front analysis. The results are in excellent agreement with those derived from the empirical functions used in section 4. Also discussed in this section is the implication on the characteristic surface wave breaking speed and breaking strength parameter. These quantities are related to the breaking intensity that is closely connected to bubble entrainment. The breaking wave speeds derived from analyses of many field measurements (Ding and Farmer 1994; Felizardo and Melville 1995; Lee et al. 1996; Terray et al. 1996; Liu et al. 1998; Frasier et al. 1998; Phillips et al. 2001; Hwang et al. 2008) also show only minor wind speed dependence (Hwang 2007, 2009). Combined with the whitecap observations, a reasonably-narrow range (within about a factor of 3) of the breaking strength parameter as a function of wind speed in open ocean conditions is suggested. Section 6 is a summary.

**2. Buoyancy of bubble plumes**

The buoyancy per unit volume of an air-water mixture with density $\rho_m$ submerged in water of density $\rho_w$ can be expressed as

$$B = -\left(\rho_m - \rho_w\right)g . \tag{1}$$

In this paper, the interest is primarily on the upper ocean surface layer relevant to whitecap observations and the water density is assumed constant. The vertical structure of the bubble plume is approximated by assigning a vertical dependence to the mixture density, that is, $\rho_m(z)$, where $z$ is the vertical coordinate with the origin at the mean water level and positive upward.

For simplicity, horizontal homogeneity is assumed within the bubble plume. The mixture density is given by

$$\rho_m = \rho_w(1-f_a) + \rho_a f_a, \tag{2}$$

where $\rho_a$ is the air density and $f_a$ is the void fraction (volume of air per unit volume of air-water mixture).

The rate of energy input per unit sea surface area required to keep the mixture submerged is approximated by

$$\begin{aligned} E_{tb} &= \sum\nolimits_{-z_b}^{0} B w_b \Delta z \\ &= \sum\nolimits_{-z_b}^{0} \rho_w (1-s) f_a g w_b \Delta z \end{aligned}, \tag{3}$$

where $s = \rho_a / \rho_w$, $z_b$ is the entrainment depth of the bubble plume, and $w_b$ is a representative terminal rise velocity of a horizontal $\Delta z$ slice of the bubble plume for buoyancy computation.

A relationship is needed between the void fraction $f_a$ and the whitecap coverage $f_w$ in order to proceed further with (3) using the whitecap observations. Ideally, we would like to have a detailed model of the depth dependence and size distribution of the bubble plume as functions of wind and wave conditions. Such information is still lacking, especially for the short duration of bubble entrainment and fragmentation when (3) can be connected to the surface wave breaking properties. Here, a simple model is presented to elicit some insight about the physical properties of the whitecap bubble plume at the initial stage of breaking entrainment. Fig. 1a depicts a bubble plume with surface area coverage $A_b$ and penetration depth $z_b$ in a water volume with surface area $A_m$ and depth $z_m$. Without loss of generality, the water volume can be assumed to be of unit surface area and unit water depth, that is, $A_m = 1$ m$^2$ and $z_m = 1$ m.

For the purpose of evaluating the buoyancy of the bubble plume, consider the cuboidal (rectangular parallelepiped) conceptual bubble plume sketched in Fig. 1b, within which all the bubbles in the bubble plume of Fig. 1a are collected and packed tightly but no interaction or interference occurs between bubbles in the conceptual bubble plume. The surface area $A_e$ of the cuboidal volume is identically $A_b$, and $z_e$ is the effective or equivalent-buoyancy depth of the bubble plume layer. That is, the buoyancy of the bubble plumes in Figs. 1a and 1b are identical. The fraction of the ocean surface covered by whitecaps is

$$f_w = \frac{A_b}{A_m} = \frac{A_e}{A_m}, \qquad (4)$$

and the void fraction of the air bubbles in the measurement volume is

$$f_a = \frac{A_e z_e}{A_m z_m} = f_w \frac{z_e}{z_m}, \qquad (5)$$

where $A_m = 1$ m$^2$ and $z_m = 1$ m are kept in the equations to keep track of the dimensions.

Substituting (5) into (3) with the summation limit $z_b$ replaced by $z_e$ and approximating $1-s \approx 1$, the solution for a bubble plume with depth-independent $f_a$ and $w_b$ becomes

$$E_{tb} = \frac{\rho_w f_w g z_e^2 w_b}{z_m}. \qquad (6)$$

The assumption of depth-independent $f_a$ and $w_b$ is a reasonable one for the conceptual bubble plume, in which well mixing of the collected bubbles is not prohibited. Alternatively, it can be applied to the special case of a bubble plume with uniform bubble size. The buoyancy formulation starting with the assumption of a bubble plume containing uniform-sized bubbles is given in Appendix A. The simpler formulation yields the same result for $E_{tb}$.

### 3. Relating bubble buoyancy to wave breaking energy dissipation

Information of the turbulence dissipation rate in the ocean, especially its vertical structure and dependence on wind and wave parameters, remains very sketchy. On the other hand, we have a better handle of the energy dissipation rate per unit ocean surface area $E_{tD}$, which is the integration of the dissipation rate per unit mass $\varepsilon$ over the water depth,

$$E_{tD} = \int_{-\infty}^{0} \rho_w \varepsilon dz. \qquad (7)$$

The expression of energy dissipation rate per unit ocean surface area is especially useful for the upper ocean layer with breaking of wind-generated surface waves as the primary source of turbulence generation. Phillips (1985) derives an analytical solution for the spectrum of energy dissipation rate of a wind-generated wave field in equilibrium state. The method has been employed in field measurements to provide useful quantitative data of the spectrally-integrated energy dissipation rate related to surface wave breaking (Felizardo and Melville 1995; Terray et al. 1996; Hanson and Phillips 1999). Hwang and Sletten (2008) show that for wind-generated surface waves, the breaking wave energy dissipation rate can be derived from the similarity relation of the wave growth function. The analysis leads to the following parameterization formula,

$$E_{tD} = \alpha \rho_a U_{10}^3, \text{ with } \alpha = 0.20 \omega_\#^{3.3} \sigma_\#, \qquad (8)$$

where $\sigma_\# = \sigma^2 g^2 / U_{10}^4$ and $\omega_\# = \omega_p U_{10} / g$ are the dimensionless wave variance and spectral peak frequency, respectively, $U_{10}$ is neutral wind speed at 10 m elevation, $\sigma$ is the root mean square wave displacement, $\omega_p$ is the spectral peak frequency, and $\rho_a = 1.2$ kg m$^{-3}$. Furthermore, the similarity relationship of wind wave growth can be expressed as

$$\sigma_\# = R\omega_\#^r, \qquad (9)$$

with the coefficients

$$R = e^{\gamma_0} \omega_\#^{-\gamma_2 \ln \omega_\#}, \quad r = \gamma_1 + 2\gamma_2 \ln \omega_\#, \qquad (10)$$

where $\gamma_0 = -6.1384$, $\gamma_1 = -2.4019$, and $\gamma_2 = -0.6102$ (e.g., Hwang et al. 2011). $E_{tD}$ in (8) thus can be given as a function of $\omega_\#$ alone, that is,

$$E_{tD} = \alpha \rho_a U_{10}^3, \text{ with } \alpha = 0.20 R \omega_\#^{3.3+r}, \quad (11)$$

where $R$ and $r$ are functions of $\omega_\#$ (10). For the wind-generated wave conditions frequently encountered in the ocean ($\omega_\#$ between about 0.8 and 3.0), the numerical value of $\alpha$ is within a narrow range between about $3.7 \times 10^{-4}$ and $5.7 \times 10^{-4}$ (Hwang and Sletten 2008, Fig. 3). The parameterization function yields very similar results as those obtained from spectral integration.

Lamarre and Melville (1991) show that a large fraction, estimated to be 30 to 50 percent and may be more, of the wave breaking energy is expended on the bubble plume buoyancy, thus we can write

$$E_{tb} = X_{Db} E_{tD}, \quad (12)$$

where $X_{Db}$ is the ratio between the rates of energy change of buoyancy and wave breaking, here $X_{Db}=0.4$ is assumed (the average of 30 and 50 percent).

With (6) and (12), a dimensionally-consistent function of the whitecap coverage can be formulated as

$$f_w = X_{Db} \frac{z_m}{z_e} \frac{E_{tD}}{\rho_w g z_e w_b}. \quad (13)$$

From (13), it becomes clear that the product of bubble properties $g z_e w_b$ forms the cubic velocity scale of the whitecap coverage $f_w$ (and $E_{tb}$), and $z_e/z_m$ relates the air volume fraction and whitecap coverage (5).

Because field observations repeatedly show that the wind speed dependence of $f_w$ and $E_{tD}$ is cubic (Fig. 2), (13) suggests that the product $z_e^2 w_b$ should be close to a constant value. This is examined in the next section. Here the empirical relationships of $f_w(U_{10})$ and $E_{tD}(U_{10})$ are discussed further. Combining (8) and (13), the familiar cubic wind speed relationship of the whitecap coverage emerges in a dimensionally-consistent formula:

$$f_w = \alpha X_{Db} \frac{\rho_a}{\rho_w} \frac{z_m}{z_e} \frac{U_{10}^3}{gz_e w_b}. \tag{14}$$

Fig. 2a shows the results of whitecap coverage from several field observations in the open ocean. The assembly of data MTRXLS (Monahan 1971; Toba and Chaen 1973; Ross and Cardone 1974; Xu et al. 2000; Lafon et al. 2004, 2007; Sugihara et al. 2007) has been described in Hwang and Sletten (2008). The more recent measurements C08 (Callaghan et al. 2008) are the results of analyzing 43158 video images using an automated whitecap extraction algorithm; each image covers an area about 4997 m$^2$. Their analysis yields 107 $f_a$ data points with wind speed coverage between 3.70 and 23.09 m/s. (To reduce clutter in the figure, the convention used in this paper for identifying the published data source in the figure legend is the initial of the lead author's last name followed by the last two digits of the publication year. If data sets from multiple publications are represented as one group, the data group is identified by the combined initials of the lead authors of the source papers.)

Fig. 2b shows the energy dissipation rate computed with the parameterization function (8) applied to several field data sets that encompass both wind-sea and mixed-sea conditions. The wind-sea category includes DMAJ (Donelan 1979; Merzi and Graf 1985; Anctil and Donelan 1996; Janssen 1997), T96 (Terray et al. 1996), D72 (DeLeonibus and Simpson 1972), and H04 (Hwang and Wang 2004a) – of which DM of the DMAJ group and T96 are under fetch-limited wave growth conditions whereas D72 and H04 are duration-limited. The mixed-sea category includes F95 (Felizardo and Melville 1995), G09 (Garcia-Nava et al. 2009) and R10 (Romero and Melville 2010). For the mixed sea data, the wave variance and peak frequency for using (8) are derived from the wind sea portion of the wave spectrum. A more detailed discussion of the wind-sea and mixed-sea wave growth functions is given in Hwang et al. (2011).

Note that the similarity relationship of wind wave growth is developed for the conditions of fully-rough surface boundary layer under turbulent wind forcing (nominally $U_{10}$ greater than about 7 m s$^{-1}$). For lower wind speeds, deviation from (8) and (14) can be expected. The cubic wind speed dependence in low wind speeds is maintained when a threshold wind speed is introduced. There are many different suggestions of the threshold wind speed in various empirical whitecap formulas, e.g., see the list compiled by Anguelova and Webster (2006). Here 2 and 3.7 m s$^{-1}$ suggested by Hwang and Sletten (2008) and Callaghan et al. (2008) are used but choosing other threshold values does not alter the results significantly. For the $z_e$ computation, $f_w$ and $E_{tD}$ are represented by the following three sets of empirical formulas:

$$\begin{aligned} f_{w0} &= 7.5 \times 10^{-6} U_{10}^3, & E_{tD0} &= 5 \times 10^{-4} \rho_a U_{10}^3, \\ f_{w2} &= 1.5 \times 10^{-5} (U_{10} - 2.0)^3, & E_{tD2} &= 1.0 \times 10^{-3} \rho_a (U_{10} - 2.0)^3, \\ f_{w3} &= 3.0 \times 10^{-5} (U_{10} - 3.7)^3, & E_{tD3} &= 2.0 \times 10^{-3} \rho_a (U_{10} - 3.7)^3. \end{aligned} \qquad (15)$$

Incidentally, several whitecap analyses separate the active and passive stages of the surface foam (e.g., Ross and Cardone 1974; Kleiss and Melville 2010). The cubic wind speed dependence is found for both types of whitecap coverage. [The more recent data of Kleiss and Melville (2010) are not included in the analysis here because they have noted (their p. 2591) that the quantitative values of their whitecap coverage are about one order of magnitude smaller than other published data. Their data are based on airborne video images, whereas those of the others are processed from ocean surface images taken at near-surface levels (from ships or ocean towers). This is an indication that the airborne measurements may have suffered from insufficient spatial resolution and optical contrast for distinguishing the smaller breakers. Black and Adams (1983) and Black et al. (1986) have assembled a catalog of sea surface images photographed from aircraft at various altitudes over a wide range of wind speeds. They have commented emphatically on the loss of surface details in the photograph with increasing aircraft

*JPO*     11     BuoyancyWhitecapR1noline.doc

altitude.]

**4. Effective rise velocity, entrainment depth, volume and surface area of entrained air bubbles**

*a. Formulation*

From (13), the effective entrainment depth can be calculated by

$$z_e = \left( \frac{X_{Db} E_{tD} z_m}{\rho_w f_w g w_b} \right)^{1/2}. \tag{16}$$

In which $X_{Db}$, $z_m$, $\rho_w$ and $g$ are constant (0.4, 1 m, 1030 kg m$^{-3}$ and 9.8 m s$^{-2}$, respectively), and empirical functions (15) are available for $f_w$ and $E_{tD}$. Specifically, the three sets of empirical functions in (15) yield $f_w = 0.0125 E_{tD}$, and the ratio between $E_{tD}$ and $f_w$ in (16) is a simple constant (0.0125) independent on wind speed.

There remain two unknowns, $z_e$ and $w_b$, in the single equation. In order to compute $z_e$, it is critical to have a good understanding of $w_b$ (the representative rise velocity of the bubbles in the bubble plume) and the associated information of the representative bubble size of the bubble plume.

*b. Bubble rise velocity and size factor*

It is well established that the terminal rise velocity of a large bubble in water departs significantly from the theoretical curve expected of rigid spheres. The deviation is caused mainly by the fluid flow within the gas bubble and the nonlinear trajectories of the rising bubble (e.g., Gaudin 1957; Clift et al. 1978; Leifer et al. 2000). The bubble rise velocity further decreases in the presence of contaminants in the water. Laboratory experiments show that the rise velocities of bubbles with radii larger than about 0.5 mm differ only slightly, as illustrated in Fig. 3, which is reproduced from Fig. 11.14 of Gaudin (1957). Similar results with more experimental data are also shown in Clift et al. (1978, Fig. 7.3) and Leifer et al. (2000, Fig. 4). Because the main interest of this paper is the initial stage of bubble plume entrained by wave breaking, it is assumed that the bubble surfaces are reasonably uncontaminated and the upper curves corresponding to $w_b(a)$ in clean or lightly contaminated water is applicable. These curves show a relatively constant rise velocity, between about 0.2 and 0.3 m s$^{-1}$, for a large bubble with radius greater than about 0.5 mm. For smaller bubbles (radii between about 0.1 and 0.3 mm), the relationship of $w_b(a)$ is approximately linear.

Deane (1997) and Deane and Stokes (1999, 2002, 2010) have presented some quantitative results of the very challenging measurements of the near-surface bubble size distribution under breaking waves. The investigations lead to the conclusion that there are two distinct mechanisms controlling the bubble size distribution under breaking waves. In laboratory measurements of bubble plumes in a sea water wave flume (Deane and Stokes 2002), they show that for bubbles with radius $a$ larger than about 1 mm, the dominant mechanism is turbulent fragmentation, which produces a bubble size distribution proportional to $a^{-10/3}$. Smaller bubbles are created by jet and drop impact on the water surface and the size distribution is proportional to $a^{-3/2}$. The two-branch size distribution of the bubble plumes under active breaking waves is also observed in the field measurements of bubble plumes 0.3 m below the water surface within a short period (order of a second) of wave breaking. The slopes of the size spectrum in the small- and large-size branches at three plume ages represented by the void fractions of 0.065, 0.027 and 0.0073 are (-1.8, -4.9), (-2.5, -5.3), and (-2.9, -5.5), respectively. The bubble radius separating the two branches is also in the neighborhood of 1 mm. The authors further indicate that the separation size scale near 1 mm is consistent with the Hinze scale of turbulent fragmentation of air bubbles (Hinze 1955) under the breaking wave conditions. The volume probability density distribution (pdf) can be expected to exhibit a prominent peak at a radius conservatively estimated to be between 0.5 and 2 mm; the slopes of the volume pdf are about (0.1 ~ 1.5) and (-0.3 ~ -2.5), respectively in the small- and large-size branches.

In a study of bubble crushing in a liquid, Longuet-Higgins (1992) gives a summary of laboratory and field measurements of bubble size distributions under wind waves, water falls and running streams. Amazingly, the distribution peaks of the bubble populations are mostly between 0.5 and 1 mm. Martínez-Bazán et al. (1999a, b) present theoretical and experimental results on the size distribution of bubbles fragmented in turbulent shear flows. Their results also show a similar range of the bubble sizes at the distribution peaks. More details of these studies are given in Appendix B. These results seem to further buttress the robustness of the peak size in the neighborhood of 1 mm for the bubble volume size distribution under active breaking waves as repeatedly observed by Deane and Stokes.

*c. Bubble entrainment computation*

As shown in Fig. 3, the rise velocity of a large bubble (radius greater than about 0.5 mm) in clean or slightly contaminated water is relatively constant. In the following $w_b$=0.25 m s$^{-1}$ is used for large bubbles. Fig. 4a presents the calculated $z_e$ using the three sets of empirical formulas of $f_w$ and $E_{tD}$ (15) described in section 3. Because $w_b$ is constant and both $E_{tD}$ and $f_w$ show cubic wind speed dependence, the resulting $z_e$ (=0.11 m) is independent on wind speed. The applicable wind speed range for this result is estimated to be between about 5 and 25 m s$^{-1}$ judging from the available field data (Fig. 2). The nearly-constant effective entrainment depth suggests strongly that the bubble entrainment process during the active wave-breaking stage is a quasi-2D problem with an almost constant vertical scale. Deane and Stokes (2002, Figs. 2 and 3) present a couple of video sequences of the bubble plume evolution that show a relatively constant bubble plume depth over a few wave cycles. The video sequences seem to offer some experimental support for the qausi-2D breaking entrainment process, as deduced from the relatively constant $z_e$ obtained from the buoyancy consideration using the conceptual bubble plume model.

The volume of air from breaking wave entrainment per unit sea surface area can be computed from the product of whitecap surface area and the effective entrainment depth (5). Because $z_e/z_m$ is essentially a constant (0.11), the formula becomes simply

$$f_a = 0.11 f_w. \tag{17}$$

Here, the water volume for $f_a$ should be restricted to the top meter of the ocean. The result of the air volume entrainment per unit sea surface area $V_{bubbles}$ is shown in the lower portion of Fig. 4a. The air volume entrainment (a 3D property) maintains the same cubic wind speed dependence as that of the whitecap area coverage (a 2D property).

We can also estimate the bubble surface area $A_{bubbles}$ of the entrained air volume by assuming a uniform bubble size. In terms of $f_w$, the volume and surface area as well as the number ($N_{bubbles}$) of bubbles entrained by breaking waves per unit sea surface area are

$$\begin{aligned} V_{bubbles} &= 0.11 f_w V_m, \\ N_{bubbles} &= 0.0825 f_w \frac{V_m}{\pi a^3}, \\ A_{bubbles} &= 0.33 f_w \frac{V_m}{a} \end{aligned} \tag{18}$$

where $V_m = A_m z_m = 1$ m$^3$, and $a$ is in meters.

Figure 4b presents three examples of the bubble surface area estimation for the effective bubble radii of 0.5, 1 and 2 mm. It is clear that the entrained bubbles provide a dramatic enhancement (orders of magnitude in high winds) of the surface area for gas or contaminant exchange. The cubic wind speed dependence is also evident in the computed surface area (a 2D property) of the entrained bubbles.

## 5. Discussion

### *a. Wind speed dependence of the effective entrainment depth*

As described in section 1, several dimensionally-consistent expressions of whitecap coverage have been developed (e.g., Toba and Chaen, 1973; Phillips 1985; Toba and Koga 1986; Zhao and Toba 2001; Yuan et al. 2009). For example, Phillips (1985) gives an analytical expression of the whitecap coverage [his (6.13)] advanced from his breaking front analysis. The solution can be written in terms of the spectrally integrated dissipation rate $\int \varepsilon(\vec{c}) d\vec{c}$, which is the equivalent of $E_{tD}/\rho_w$ in this paper, and $\vec{c}$ is the phase velocity vector of the breaking front. Using his (6.6) for $\varepsilon(\vec{c})$, then

$$f_w = \frac{1}{4b \ln(c_{max}/c_{min})} \frac{gT_b}{c_{min}^4} \frac{E_{tD}}{\rho_w}, \qquad (19)$$

where $b$ is a numerical constant (breaking strength parameter) in the calculation of the rate of energy loss per unit length of the breaking front, $T_b$ is the average bubble persistence time on the water surface after generation, and $c_{min}$ and $c_{max}$ are the lower and upper integration limits of the phase velocity of wave breaking fronts.

The $c_{min}$ can be considered as the lower threshold of the breaking front phase speed that is capable of producing whitecaps (Phillips 1985), with its quartic dependence, it is the most sensitive one among the group of "free parameters" in (19) for quantitative evaluation. Field measurements suggest that the range of $c_{max}/c_{min}$ is about one order of magnitude, and $c_{min}$ is in the neighborhood of about 1 m s$^{-1}$ (e.g., Frasier et al. 1998; Melville and Matusov 2002; Hwang et al. 2008). The numerical value of $b$ is estimated to be about 0.06 by Phillips (1985) based on the results from a series of laboratory experiments by Duncan (1981) but later analyses indicate that $b$ may vary over a wide range. Further discussion on this breaking strength parameter is deferred to section 5*c*. Zheng et al. (1983) measure the life time of bubbles on the still water surface. The bubbles are generated by forcing pressurized air to exit a capillary tube positioned at 0.1 m below the water surface. The bubble size is controlled by the nozzle size of the capillary tube. Bubbles with radii ranging from 0.7 to 3.7 mm are produced with different capillary tubes. There is a general trend of increasing life time with the bubble size, reaching a maximum near 2 to 3 mm radius, and then the life time decreases for larger bubbles. The overall average life times for various bubble sizes are 2.24, 2.98, and 3.89 s for tap water, Delaware Bay water and Atlantic Ocean water, respectively. It is not clear about the effects on the bubble life time by turbulence and other disturbances induced by wind or water currents.

Figure 5 shows the MTRXLS whitecap observations that include wave data for the calculation of the wave breaking energy dissipation rate using (8), more details are given in Hwang and Sletten (2008). For comparison, the results of $f_w(E_{tD})$ obtained by Phillips (1985) solution (19) with $b$=0.06, $T_b$=3 s, $c_{max}/c_{min}$=10, and $c_{min}$ between 1 and 2 m s$^{-1}$ are superimposed on the figure (labeled as the P85 curves). The three sets of empirical functions in (15) yield $f_w$=0.0125$E_{tD}$, which is in close agreement with (19) for $c_{min}$=1.4 m s$^{-1}$, these two curves go through the middle of the data cloud.

Based on the comparison with data and given the present state of imperfect knowledge about the values of $b$, $T_b$, $c_{max}$, and $c_{min}$, it seems that in (19), treating the most sensitive factor $c_{min}$ as the threshold phase speed (about 1 m s$^{-1}$) of whitecap generation would produce an upper bound of whitecap observations. It can also be interpreted as the characteristic breaking front phase speed (about 1.4 m s$^{-1}$) to represent the mean trend of whitecap measurements. The factor $c_{max}/c_{min}$ remains the ratio of the upper and lower bounds of the integration range, but its impact is relatively small due to the logarithmic dependence. Note that there is a wide range of $b$ suggested in the literature, ranging from $7\times10^{-5}$ to $7.5\times10^{-2}$, as shown in Figs. 2 and 16 of Drazen et al. (2008). The effect of varying $b$ on the interpretation of the characteristic breaking wave speed as represented by $c_{\min}$ is deferred to section 5c.

Equating (13) with (19), an independent estimate of the effective entrainment depth can be obtained from

$$z_e = \left(4bX_{Db}\ln(c_{\max}/c_{\min})\frac{z_m c_{\min}^4}{g^2 T_b w_b}\right)^{1/2}. \tag{20}$$

The result, similar to (16), again shows that the effective bubble plume entrainment depth is independent or at most weakly dependent on wind speed. The numerical value of $z_e$ from (20) is also 0.11 m for $X_{Db}=0.4$, $b=0.06$, $T_b=3$ s, $c_{\max}/c_{\min}=10$, and $c_{\min}=1.4$ m s$^{-1}$.

Interestingly, if $z_e$ is not constant but depends on wind speed to some power, then the dependence on wind speed for the volume and surface area of the entrained bubbles (section 4c) would differ from cubic for the situation that $f_w$ increases cubically with wind speed (5). The volume of entrained air is directly proportional to the bubble plume buoyancy that represents a big portion of the wave breaking energy dissipation. Given that wind speed cubed is such a strong signature of wave breaking properties (Phillips 1985, 1988), the dependence on wind speed in $z_e$ is not likely to be strong.

*b. Characteristic breaking wave speed*

One measure of the breaking intensity in connection to bubble entrainment is the breaking wave speed. Many field measurements of the of breaking event speeds have been obtained using acoustic noise event tracking (Ding and Farmer 1994), Doppler or feature tracking of radar sea spikes (Lee et al. 1996; Frasier et al. 1998; Liu et al. 1998; Phillips et al. 2001; Hwang et al. 2008), video tracking of whitecap evolution (Melville and Matusov 2002; Kleiss and Melville 2010), and source function balance analysis of the wave action conservation equation (Hwang and Wang 2004b). Hwang (2007, Fig. 4) presents the breaking wavelength (converted from wave speed) as a function of wind speed with data obtained from all the different approaches described above. The source function balance analysis of Hwang and Wang (2004) is applied to short scale waves measured by free-drifting wave gauges that effectively high-pass filtered the wave signals. The airborne video images of whitecaps may have resolution and contrast issues for resolving smaller whitecap patches, as has been discussed in section 3. The results of breaking wave speed from acoustic and radar measurements are illustrated in Fig. 6. Remote sensing techniques usually process a very large population of breaking events, on the order of tens of thousands.

Hwang (2009) describes an approach using empirical formulations to obtain the energy transfer velocity across the air-sea interface based on the ratio between the breaking wave energy dissipation rate and surface wind stress (Gemmrich et al. 1994; Terray et al. 1996). The energy transfer velocity is found to be in close agreement with the breaking wave speed. The results of energy transfer velocity processed by Hwang (2009) using the wind and wave measurements of young wind seas in a lake by Terray et al. (1996) and more mature wind seas mixed with swell in deep ocean by Felizardo and Melville (1995) are also shown in Fig. 6. It is noteworthy that for each individual data set, the obtained breaking wave speeds vary within a small range especially in higher winds, and they display only a weak dependence on wind speed. Combined together, although many of the data points are grouped quite tightly, the data sets of Ding and Farmer (1994) and Lee et al. (1996) show apparent departure from the majority of the data groups. As pointed out by Hwang (2007):

"Two major factors contribute to the observed wide range of the breaking length scales at a given wind speed. The first is the dynamic range of the sensors. In particular, the passive acoustic tracking of breaking events cannot detect small-scale breakings due to the problem of ambient noise (Ding and Farmer, 1994). The second factor is the stage of wave development. The experiments reported by Lee et al. (1996) are conducted in a lake and a protected bay with limited wind fetch so the wave field is relatively young compared to the wave conditions of the others obtained in the open ocean environment."

Taking these two factors into consideration, the two lake data sets of young sea conditions (L96 and T96) and the acoustic measurements (D94) are excluded in the subsequent analysis. The breaking wave speeds of open ocean waves are shown to distribute within a narrow range between about 2 and 3.5 m s$^{-1}$ for wind speeds ranging between about 5 and 16 m s$^{-1}$. The variation is especially small in higher wind speeds (for $U_{10}$ greater than about 7 m s$^{-1}$, $c_b$ is mostly within 20% of 2.8 m s$^{-1}$). The collection of the open ocean data can be fitted with the following empirical formula:

$$c_b = \min\left[0.12(U_{10}-5)+2.0, 2.8\right]. \tag{21}$$

The narrow range of the measured breaking wave speed in the ocean is in accord with the nearly-constant effective bubble entrainment depth derived from the bubble plume buoyancy consideration. Combining these two pieces of information together, a picture emerges about the wind dependence of surface wave breaking showing that the lateral breaking surface area, and probably the breaking frequency, increases at a rate proportional to wind speed cubed while the vertical reach of the bubble plume maintains a relatively constant depth during the active breaking stage. Of course, the subsequent entrainment and distribution of the bubble plume beyond the initial active breaking stage may be influenced by the wind condition because the long-term resultant turbulent and coherent flow patterns in the upper ocean layer are controlled by winds and waves. However, in the absence of other sinks and sources, the volume of the entrained air should remain the same as that entrained during the active breaking stage.

As a final comment on the breaking wave speed, although the range of available wind speeds in Fig. 6 is limited (less than 16 m s$^{-1}$), the empirical equation of the breaking wave speed (21) can probably extend to about 20 m s$^{-1}$ or higher winds. This is deduced from the bubble entrainment analysis presented in this paper. The highest wind speeds available for the analysis are about 24 and 20 m s$^{-1}$, respectively in the data of whitecaps and wave energy dissipation rate (Fig. 2).

*c. Estimating the breaking strength parameter b from whitecap observations*

As noted earlier in section 5*a*, reported values of Phillips's breaking strength parameter *b* in the literature cover a wide range of about two orders of magnitude from $7\times10^{-4}$ to $7.5\times10^{-2}$ (Drazen et al. 2008, Fig. 2). The wide range reflects the various approaches of simulating wave breaking in laboratory (e.g., steady vs. unsteady breaking, plunging vs. spilling) and the resulting breaking strength. This range expands to three orders of magnitude from $7\times10^{-5}$ to $7.5\times10^{-2}$ when the experiment results designed for near-threshold (incipient) breaking conditions are included (Drazen et al. 2008, Fig. 16).

In section 5*a*, $c_{min}=1.4$ m s$^{-1}$ is found to be in best agreement with the whitecap observations using *b*=0.06, which is near the upper bound of the *b* values (based on analysis of steady breaking laboratory experiments) cited in Drazen et al. (2008). It is also commented that $c_{min}$ can be interpreted as the characteristic breaking front phase speed to represent the mean trend of whitecap measurements. Because of the $bc_{min}^4$ combination in (19), we can quantify the effect of choosing different *b* on $c_{min}$. In particular, doubling the value of $c_{min}$ to 2.8 m s$^{-1}$ is equivalent to choosing *b*=0.06/16=$3.75\times10^{-3}$, which falls near the lower bound of the *b* range cited in Drazen et al. (2008, Fig. 2); the lower bound values are generally associated with field observations.

Alternatively, making use of the whitecap observations we can use the breaking velocity measurements to infer the value of $b$ by substituting $c_b$ for $c_{min}$ in (20). This is represented by

$$b = \frac{1}{4bX_{Db}\ln(c_{max}/c_{min})} \frac{g^2 z_e^2 T_b w_b}{z_m c_b^4}. \quad (22)$$

As commented in section 5a, there remain some uncertainties on the values of $T_b$ and $c_{max}/c_{min}$. Fig. 7a shows the calculated $b$ for a range of $c_b$ with $(T_b, c_{max}/c_{min})$=(2, 10) and (4, 5). For $c_b$ between 2 and 3.5 m s$^{-1}$, which approximately envelop the open ocean measurements discussed in section 5b, the $b$ value is between $1\times10^{-3}$ and $3\times10^{-2}$. Using the empirical formula (21) for the mean trend of the breaking wave velocity as a function of wind speed, the dependence of $b(U_{10})$ is illustrated in Fig. 7b for $(T_b, c_{max}/c_{min})$= (2, 10) and (4, 5).

## 6. Summary

The close connection between whitecap coverage and surface wave breaking has been well established from decades of field observations. Experimental evidence also indicates that a large portion of the wave breaking energy dissipation is expended to counter the buoyancy of the bubble plume. A conceptual model is developed to make use of these observations for extracting information of the air volume entrained by breaking waves and for estimating the dramatic enhancement of the air-water interface area of the entrained bubbles. During the process, it is found that, in terms of the bubble properties, the source of the cubic velocity scale in the whitecap empirical relationship is the product $gz_e w_b$. For large bubbles in the initial bubble plume, $w_b$ is almost independent on the bubble size and $z_e$ is relatively constant (about 0.11 m) and independent on wind speed. The relationship between the void fraction of the top meter ocean layer and whitecap coverage can be approximated by $f_a=0.11f_w$.

With a nearly-constant $z_e$, the initial stage of the bubble plume evolution thus becomes a

quasi-2D process. High speed video recordings of the bubble plume evolution in a sea water wave flume (Deane and Stokes 2002, Figs. 2 and 3), showing nearly-constant bubble plume depth over several wave cycles, seem to support this conclusion. In related measurements of wave breaking properties, the breaking wave speed is found to depend on wind speed only weakly, and the magnitude is within a narrow range between about 2 and 3.5 m s$^{-1}$ over a wide range of wind speeds in open ocean conditions. The relatively invariable breaking wave speed is consistent with the nearly-constant effective bubble entrainment depth from bubble plume buoyancy consideration. As illustrated in section 5*c*, whitecap observations can also be used to infer some key parameters relevant to breaking wave energy dissipation, such as Phillips's breaking strength parameter *b*, which has been poorly constrained so far and the reported values vary over about three orders of magnitude (e.g., see the summary by Drazen et al. 2008). The wide range of the reported values for *b* is attributed largely to the simulated breaking conditions in the laboratory experiments, such as steady vs. unsteady breaking, plunging vs. spilling breaking, and incipient vs. more mature stage of the breaking process. Using the field observations of whitecaps and empirical relation of the breaking wave energy dissipation rate derived from the similarity relation of ocean wind wave growth, the wind speed dependence of the average *b* for open ocean condition is illustrated in Fig. 7b, which displays a much tighter variability envelop (about a factor of 3).

    The results from the present analysis may lead to a better understanding of the surface wave breaking mechanism. They may also provide useful quantitative information for the formulation of source functions of various oceanographic processes such as gas or contaminant exchange and acoustic noise generation in the upper ocean layer.

**Acknowledgements**

This work is sponsored by the Office of Naval Research (NRL program element 61153N). Discussions with M. D. Anguelova have led to the comparison study with Phillips (1985) whitecap function described in section *5a* and improvement in the overall presentation. A. Callaghan generously provided the tabulated whitecap data set (C08 shown in Fig. 2a) with the corresponding environmental conditions. The whitecap images for these data were collected by G. de Leeuw and L. Cohen during the Marine Aerosol Production (MAP) campaign headed by C. O'Dowd and G. de Leeuw.

**Appendix A. Alternative derivation of bubble plume buoyancy and rate of energy change**

The buoyancy force of a bubble of volume $V_1$ can be written as

$$B_1 = -(\rho_a - \rho_w)gV_1. \tag{A1}$$

The rate of energy change of the bubble with rise velocity $w_b$ caused by the buoyancy is

$$e_{tb1} = B_1 w_b. \tag{A2}$$

For $N$ non-interacting uniform-sized bubbles in a unit volume of water, the total buoyancy is simply $B_N = NB_1$ and the rate of energy change is $e_{tbN} = Ne_{tb1}$, thus the buoyancy per unit volume is

$$B_N = -(\rho_a - \rho_w)gf_a \approx \rho_w g f_a, \tag{A3}$$

where $f_a$ is substituted for $NV_1/1\text{m}^3$, which is the void fraction by definition, and $s = \rho_a/\rho_w \ll 1$ is applied. Similarly, the rate of energy change per unit volume due to bubble buoyancy is

$$e_{tbN} = B_N w_b, \tag{A4}$$

With reference to Fig. 1 and (5), we can estimate the buoyancy-induced rate of energy change per unit ocean surface area by

$$E_{tb} = \sum_{-z_b}^{0} e_{tbN} \Delta z \approx \rho_w f_a g z_e w_b = \frac{\rho_w f_w g z_e^2 w_b}{z_m}, \tag{A5}$$

which is identical to (6).

**Appendix B. A brief review of bubble fragmentation in turbulent shear flows**

Martínez-Bazán et al. (1999a, b) present theoretical and experimental results on the bubble fragmentation in turbulent shear flows. One of the most interesting results relevant to this study is their finding of the pdf of daughter bubbles showing a pronounced peak at a size that is about 80% of the mother bubble size, which corresponds to the situation that the mother bubble is most likely breaking into two daughter bubbles of equal volume [$0.8 \approx (1/2)^{1/3}$]. At very high dissipation rates, tertiary splitting may become dominant and the peak of the size spectrum shifts to 70% [$0.7 \approx (1/3)^{1/3}$] of the mother bubble size (Martínez-Bazán et al. 1999b, Figs. 3 and 9). The paper by Martínez-Bazán et al. (1999b) includes an appendix by Longuet-Higgins comparing his own theoretical model of crushing air cavities in a liquid (Longuet-Higgins 1992) with two sets of the experimental data of Martínez-Bazán et al. (1999a) that provide the necessary information for the computation. The experimental data are in good agreement with the Longuet-Higgins crushing model when the average numbers of daughter bubbles are about 2.6 and 4.5 for the two cases. The corresponding size ratio of the daughter and mother bubbles are 0.73 and 0.61, respectively. So, for an original bubble plume with some peaked size distribution, the fragmented bubble plume may maintain a similar size distribution to that of the original bubble plume with the peak shifted to a slightly smaller size; the ratio of the two peak sizes is very likely to be within a factor of two.

As mentioned in section 4, Longuet-Higgins (1992) summarizes the results of several laboratory and field measurements of bubble size distributions under wind waves, water falls and running streams. Amazingly, the bubble population distribution peaks are mostly between 0.5 and 1 mm. These results seem to further buttress the robustness of the bubble volume distribution peak size in the neighborhood of 1 mm under active breaking waves as repeatedly observed by Deane and Stokes.

limited seas. *J. Phys. Oceanogr.*, **40**, 2575-2604.

Lafon, C., J. Piazzola, P. Forget, O. Le Calve, and S. Despiau, 2004: Analysis of the variations of the whitecap fraction as measured in a coastal zone. *Bound.-Layer Meteorol.*, **111**, 339-360.

Lafon, C., J. Piazzola, P. Forget, and S. Despiau, 2007: Whitecap coverage in coastal environment for steady and unsteady wave field conditions. *J. Mar. Syst.*, **66**, 38-46.

Lamarre, E., and W. K. Melville, 1991: Air entrainment and dissipation in breaking waves. *Nature*, **351**, 469-472.

Lee, P. H. Y., J. D. Barter, E. Caponi, M. Caponi, C. L. Hindman, B. M. Lake, and H. Rungaldier, 1996: Wind-speed dependence of small-grazing-angle microwave backscatter from sea surfaces. *IEEE Trans. Antenna Propagat.*, **44**, 333-340.

Leifer, I., R. K. Patro, and P. Bower, 2000: A study on the temperature variation of rise velocity for large clean bubbles. *J. Atmos. Oceanic Tech.*, **17**, 1392-1402.

Liu, Y., S. J. Frasier, and R. E. McIntosh, 1998: Measurement and classification of low-grazing-angle radar sea spikes. *IEEE Trans. Antenna Propagat.*, **46**, 27-40.

Longuet-Higgins, M., 1992: The crushing of air cavities in a liquid. *Proc. R. Soc. Lond. A,* **439**, 611-626.

Martínez-Bazán, C., J. L. Montañés, and J. C. Lasheras, 1999a: On the breakup of an air bubble injected into a fully developed turbulent flow. Part 1. Breakup frequency. *J. Fluid Mech.*, **401**, 157–182.

Martínez-Bazán, C., J. L. Montañés, and J. C. Lasheras, 1999b: On the breakup of an air bubble injected into a fully developed turbulent flow. Part 2. Size PDF of the resulting daughter bubbles. *J. Fluid Mech.*, **401**, 183–207.

Martínez-Bazán, C., J. Rodríguez- Rodríguez, G. B. Deane, J. L. Montañés, and J. C. Lasheras,
*JPO*     30     BuoyancyWhitecapR1noline.doc

**List of Figures**

Fig. 1. A conceptual sketch for the consideration of bubble plume buoyancy: (a) a simplified bubble plume, and (b) the conceptual bubble plume with identical ocean surface coverage and buoyancy as those of the bubble plume depicted in (a).

Fig. 2. (a) Field observations of whitecap coverage (MTRXLS and C08) and examples of three empirical functions of cubic wind speed dependence. (b) Same as (a) but for the breaking wave energy dissipation rate calculated with the wind wave growth function (Hwang and Sletten 2008). Field data used in the calculation include fetch-limited wind seas (DM of DMAJ and T96), duration-limited wind seas (H04 and D72), and mixed seas (F95, G09 and R10).

Fig. 3. Terminal velocities of air bubbles in distilled water and contaminated water, reproduced from Fig. 11.14 of Gaudin (1957). Similar results are also reported by Clift et al. (1978, Fig. 7.3) and Leifer et al. (2000, Fig. 4).

Fig. 4. (a) Effective entrainment depth and volume of air entrainment in the bubble plume per unit ocean surface area, (b) the surface area of entrained bubbles per unit ocean surface area estimated with the assumption of uniform bubble size.

Fig. 5. Field observations of whitecap coverage (MTRXLS) and its linear dependence on the breaking wave energy dissipation rate. The three sets of empirical functions in (15) yield $f_w=0.0125E_{tD}$, and Phillips (1985) solutions (labeled P85) computed with three different assumptions of $c_{min}$ in (19) are illustrated for comparison.

Fig. 6. Characteristic breaking wave speed represented by the breaking event speed derived from radar (L96, F98, P01, and H08) and acoustic (D94) processing, as well as the energy transfer velocity computed from the ratio between breaking wave energy dissipation rate and surface

wind stress (F95 and T96 data, processed in H09). Reasons for the departure of D94 (instrument sensitivity) and L96 (young wave condition) data sets from the others are discussed in the text. The empirical fitting curve (21) is shown with the solid line.

Fig. 7. Estimation of the breaking strength parameter $b$ combining the empirical results of whitecap observations and breaking wave energy dissipation function applied to the Phillips (1985) whitecap function. Shown here are: (a) $b$ as a function of $c_b$, the frequently observed range of $c_b$ between 2 and 3.5 m s$^{-1}$ in the open ocean data for wind speed higher than 5 m s$^{-1}$ (Fig. 6) are marked with vertical dashed-dotted lines; and (b) $b$ as a function of $U_{10}$ through the application of (21).

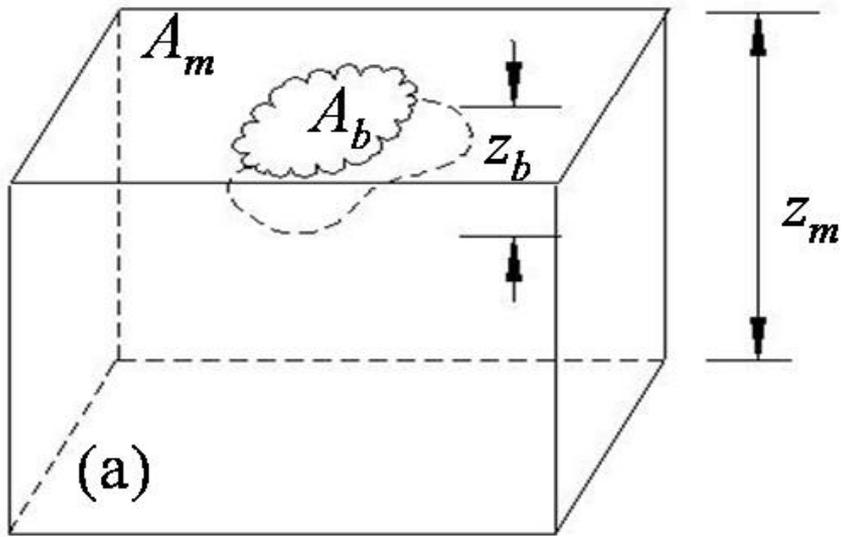 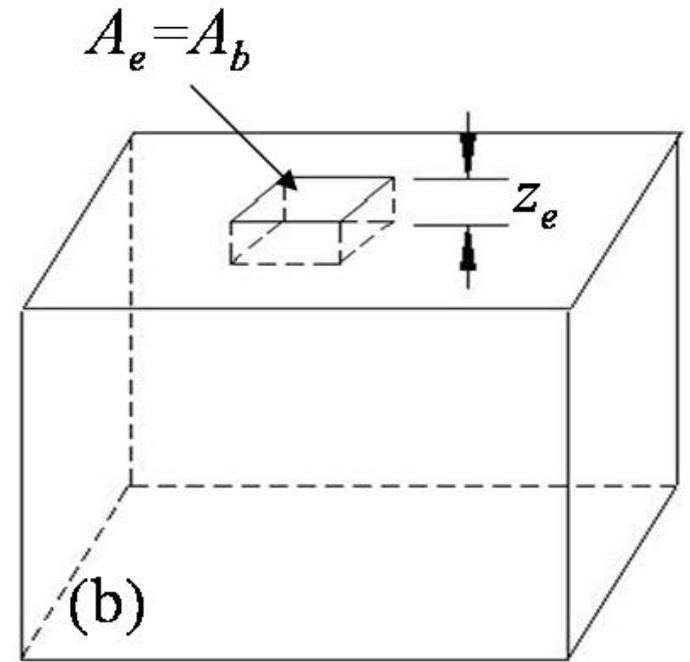

Fig. 1. A conceptual sketch for the consideration of bubble plume buoyancy: (a) a simplified bubble plume, and (b) the conceptual bubble plume with identical ocean surface coverage and buoyancy as those of the bubble plume depicted in (a).

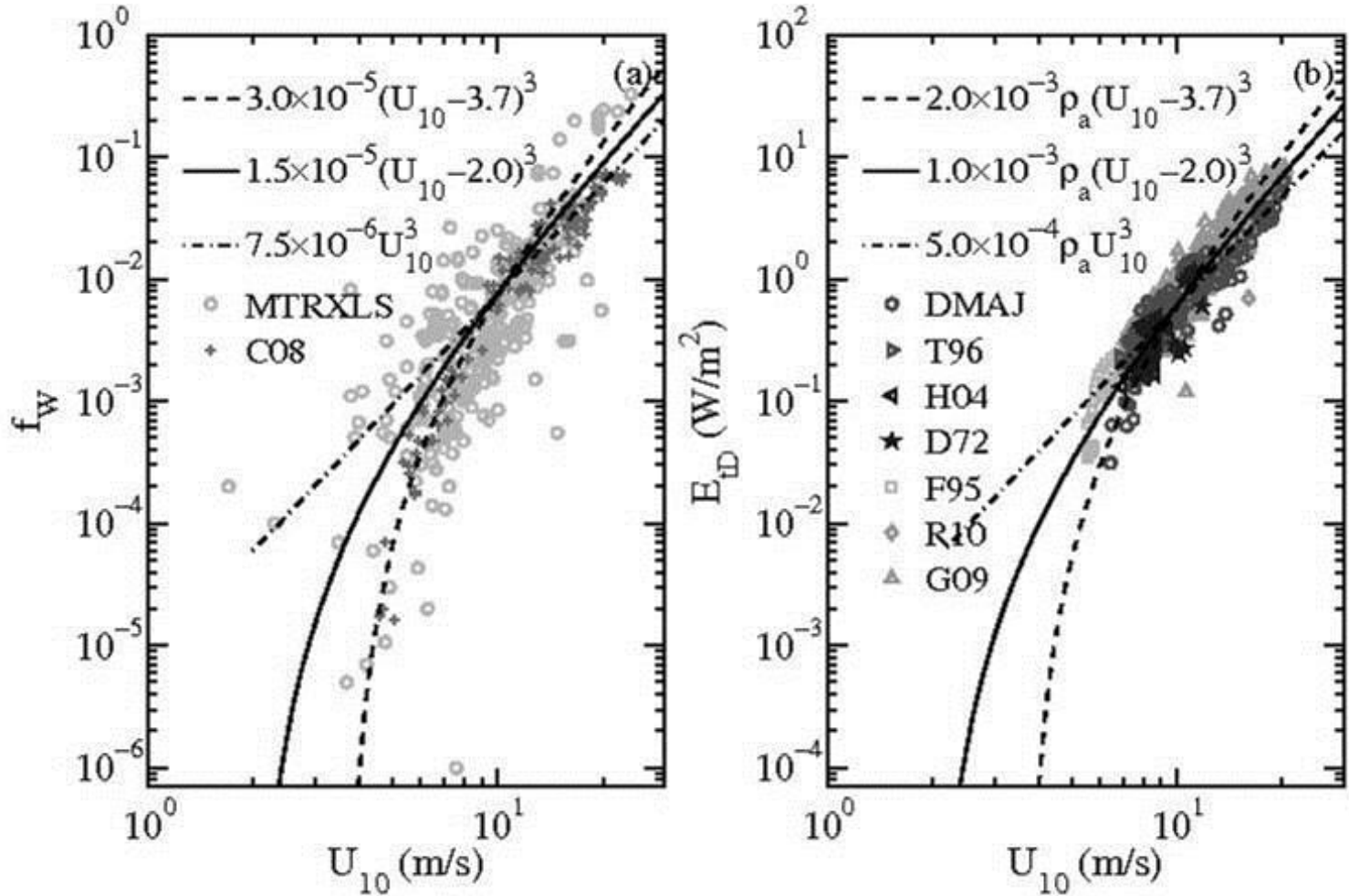

Fig. 2. (a) Field observations of whitecap coverage (MTRXLS and C08) and examples of three empirical functions of cubic wind speed dependence. (b) Same as (a) but for the breaking wave energy dissipation rate calculated with the wind wave growth function (Hwang and Sletten 2008). Field data used in the calculation include fetch-limited wind seas (DM of DMAJ and T96), duration-limited wind seas (H04 and D72), and mixed seas (F95, G09 and R10).

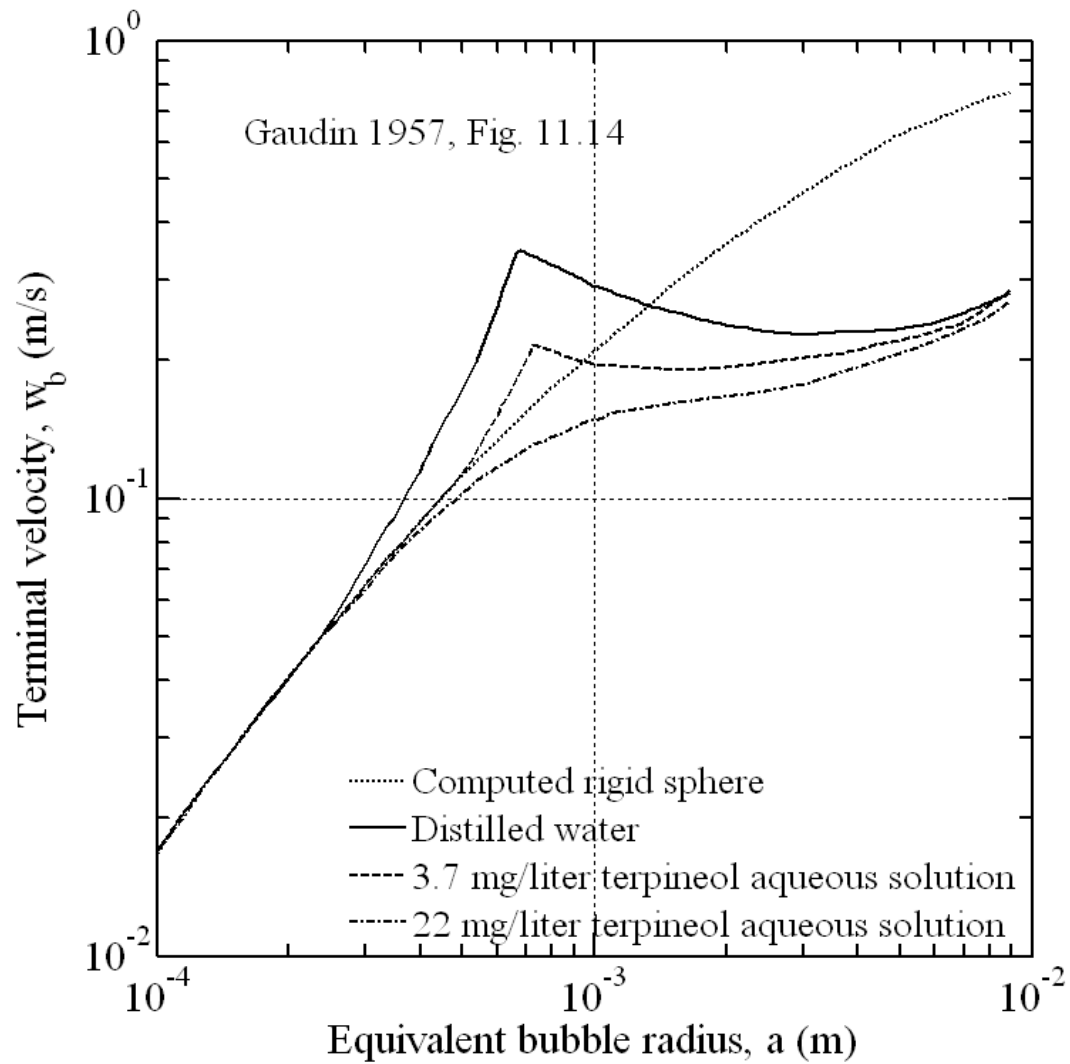

Fig. 3. Terminal velocities of air bubbles in distilled water and contaminated water, reproduced from Fig. 11.14 of Gaudin (1957). Similar results are also reported by Clift et al. (1978, Fig. 7.3) and Leifer et al. (2000, Fig. 4).

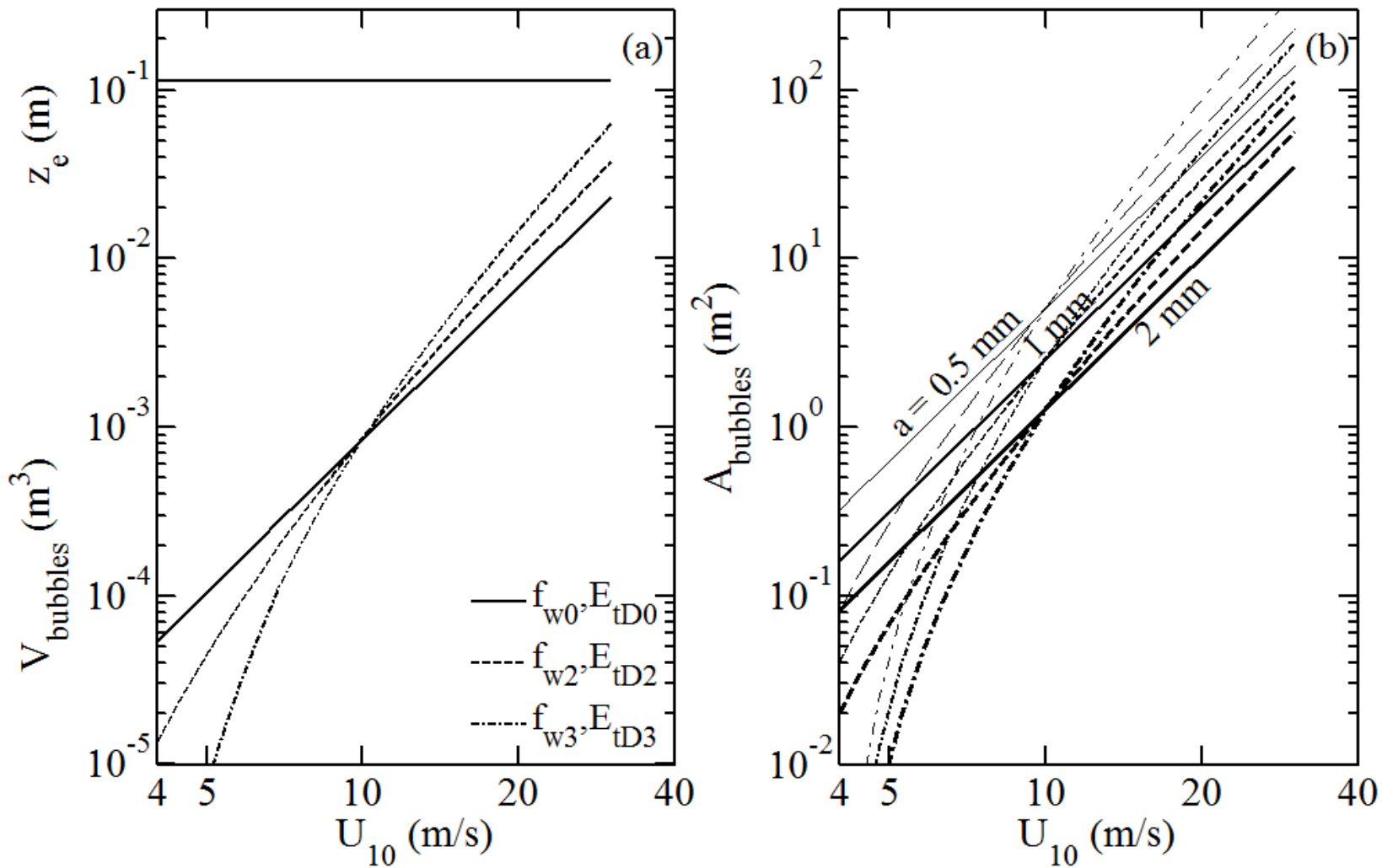

Fig. 4. (a) Effective entrainment depth and volume of air entrainment in the bubble plume per unit ocean surface area, (b) the surface area of entrained bubbles per unit ocean surface area estimated with the assumption of uniform bubble size.

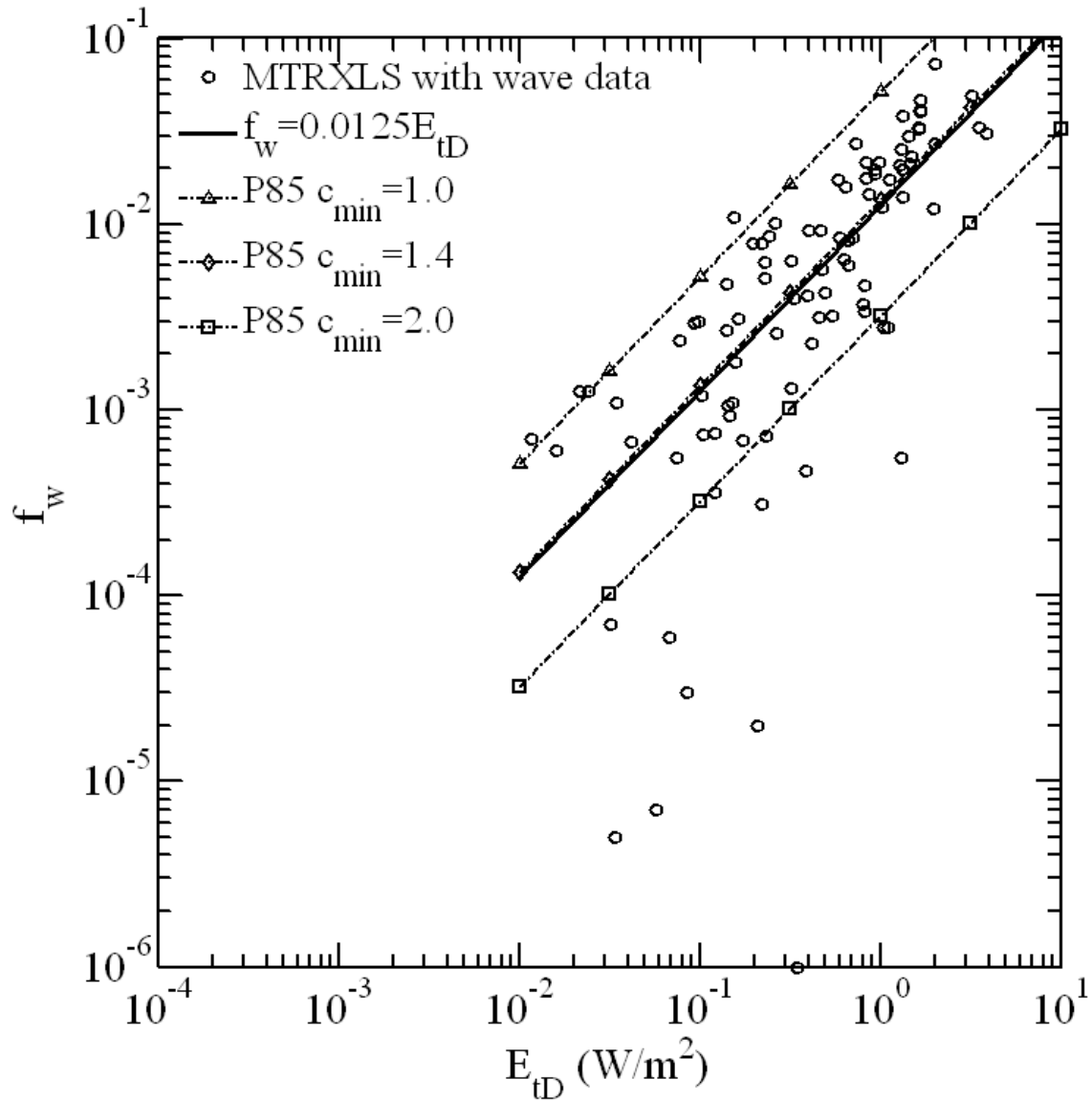

Fig. 5. Field observations of whitecap coverage (MTRXLS) and its linear dependence on the breaking wave energy dissipation rate. The three sets of empirical functions in (15) yield $f_w=0.0125E_{tD}$, and Phillips (1985) solutions (labeled P85) computed with three different assumptions of $c_{min}$ in (19) are illustrated for comparison.

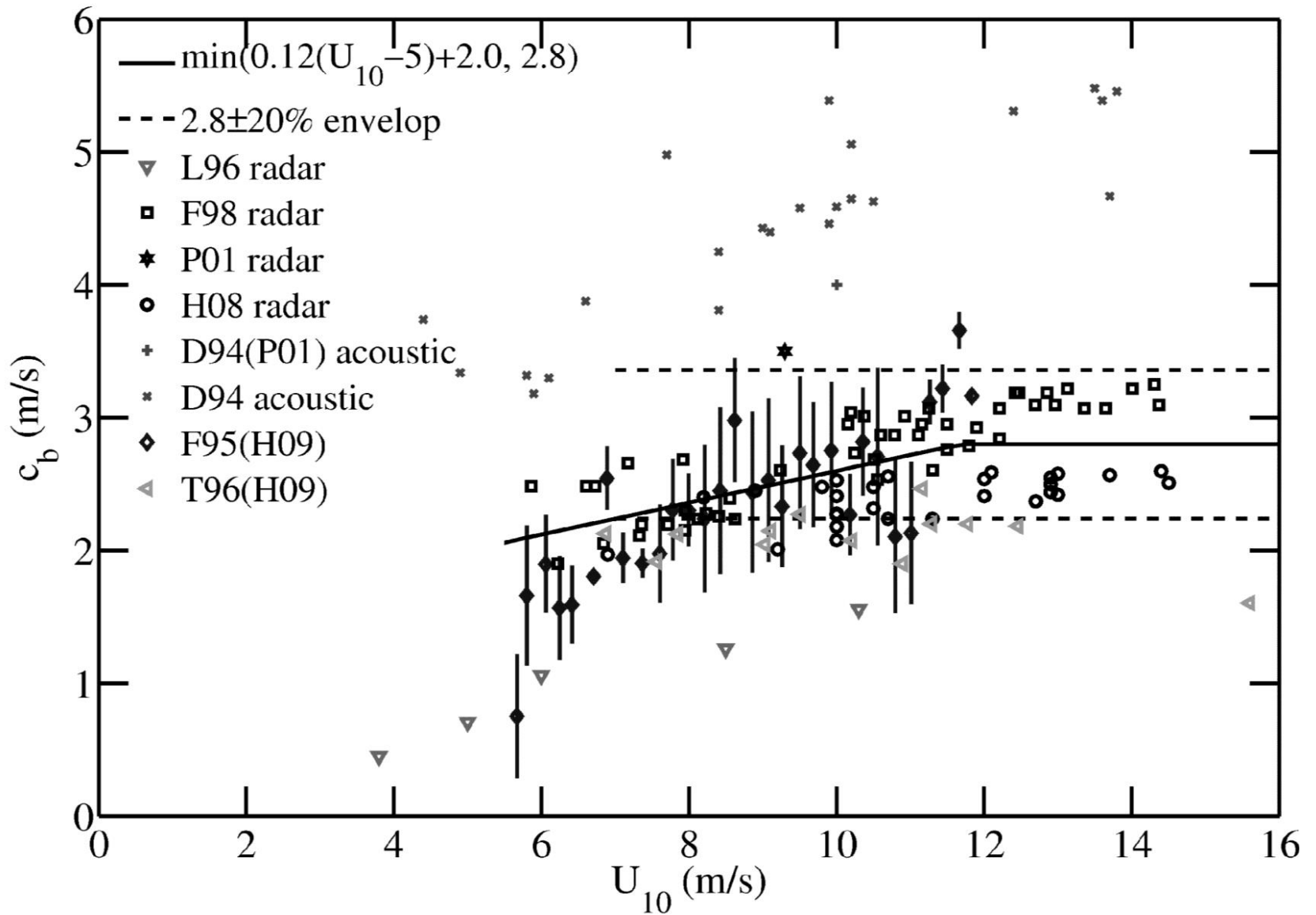

Fig. 6. Characteristic breaking wave speed represented by the breaking event speed derived from radar (L96, F98, P01, and H08) and acoustic (D94) processing, as well as the energy transfer velocity computed from the ratio between breaking wave energy dissipation rate and surface wind stress (F95 and T96 data, processed in H09). Reasons for the departure of D94 (instrument sensitivity) and L96 (young wave condition) data sets from the others are discussed in the text. The empirical fitting curve (21) is shown with the solid line.

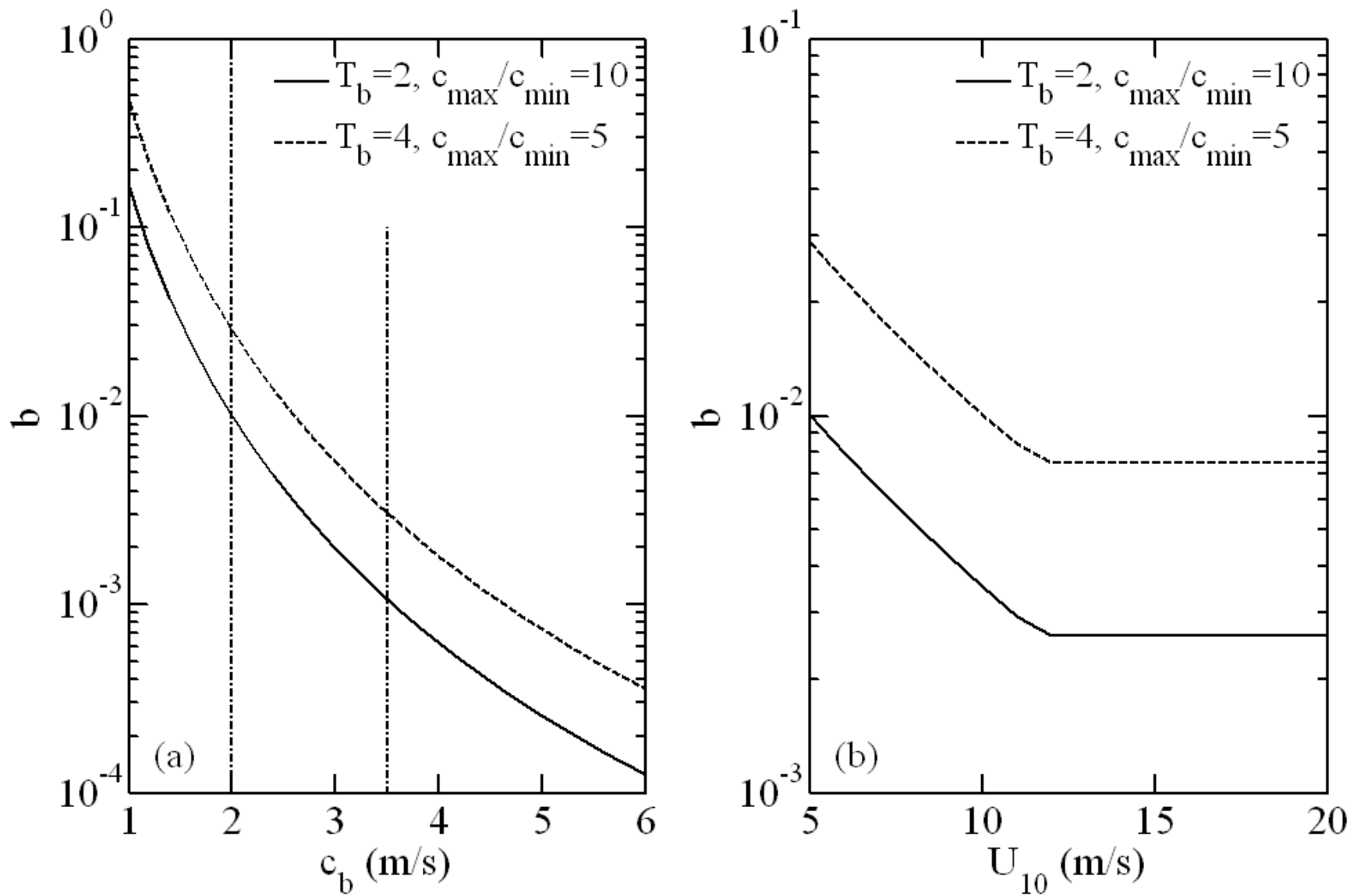

Fig. 7. Estimation of the breaking strength parameter $b$ combining the empirical results of whitecap observations and breaking wave energy dissipation function applied to the Phillips (1985) whitecap function. Shown here are: (a) $b$ as a function of $c_b$, the frequently observed range of $c_b$ between 2 and 3.5 m s$^{-1}$ in the open ocean data for wind speed higher than 5 m s$^{-1}$ (Fig. 6) are marked with vertical dashed-dotted lines; and (b) $b$ as a function of $U_{10}$ through the application of (21).